# Unleashing the Power of AI. A Systematic Review of Cutting-Edge Techniques in AI-Enhanced Scientometrics, Webometrics, and Bibliometrics

Hamid Reza Saeidnia
Department of Information Science and Knowledge Studies,
Tarbiat Modares University, Tehran, Islamic Republic of Iran

Elaheh Hosseini
Department of Information Science and Knowledge Studies,
Faculty of Psychology and Educational Sciences, Alzahra University,
Tehran, Islamic Republic of Iran

Shadi Abdoli
Department of Information Science, Université de Montreal, Montreal, Canada

Marcel Ausloos
School of Business, University of Leicester, Leicester, UK
and
Bucharest University of Economic Studies, Bucharest, Romania

**Abstract**

**Purpose:** The study aims to analyze the synergy of Artificial Intelligence (AI), with scientometrics, webometrics, and bibliometrics to unlock and to emphasize the potential of the applications and benefits of AI algorithms in these fields.

**Design/methodology/approach:** By conducting a systematic literature review, our aim is to explore the potential of AI in revolutionizing the methods used to measure and analyze scholarly communication, identify emerging research trends, and evaluate the impact of scientific publications. To achieve this, we implemented a comprehensive search strategy across reputable databases such as ProQuest, IEEE Explore, EBSCO, Web of Science, and Scopus. Our search encompassed articles published from January 1, 2000, to September 2022, resulting in a thorough review of 61 relevant articles.

**Findings: (i)** Regarding scientometrics, the application of AI yields various distinct advantages, such as conducting analyses of publications, citations, research impact prediction, collaboration, research trend analysis, and knowledge mapping, in a more objective and reliable framework. **(ii)** In terms of webometrics, AI algorithms are able to enhance web crawling and data collection, web link analysis, web content analysis, social media analysis, web impact analysis, and recommender systems. **(iii)** Moreover, automation of data collection, analysis of citations, disambiguation of authors, analysis of co-authorship networks, assessment of research impact, text mining, and recommender systems are considered as the potential of AI integration in the field of bibliometrics.

**Originality/value:** This study covers the particularly new benefits and potential of AI-enhanced scientometrics, webometrics, and bibliometrics to highlight the significant prospects of the synergy of this integration through AI.

**Introduction**

Artificial Intelligence (AI) has revolutionized various fields, in particular scientometrics, webometrics, and bibliometrics [1, 2]. Scientometrics is a field that involves the quantitative analysis of scientific literature to measure various aspects of scientific research, such as productivity, impact, and collaboration patterns [3]. It uses bibliographic data and citation analysis to understand the dynamics of scientific knowledge production and dissemination [4].

Webometrics, on the other hand, focuses on the quantitative analysis of web-based information, particularly websites and hyperlinks, to assess the impact and visibility of individuals, organizations, or research institutions on the web [5]. It employs web crawling and link analysis techniques to examine web-based structures and interactions [6].

Bibliometrics is a field that applies mathematical and statistical methods to analyze patterns of publication, citation, and collaboration in academic literature [7]. It measures the impact and influence of scholarly publications, authors, and institutions based on citation analysis and other bibliographic data [8].

These three fields are closely related to each other as they all involve the quantitative analysis of information and aim to provide insights into the production, dissemination, and impact of scientific knowledge. They share common methodologies and techniques, such as data mining, network analysis, and statistical modeling.

In the following, we demonstrate prospects based on previous applications. Furthermore, we conclude that we provide also ground for further research and prospective innovation in the field of informetrics, ultimately leading to more accurate, efficient, and insightful analyses in evidence-based decision-making.

Researchers face a challenge when dealing with the availability of vast amounts of scholarly publications, as it becomes difficult to extract knowledge, improve data analysis, and make well-informed decisions. AI-enhanced algorithms and techniques have played a crucial role in automating the identification, classification, and analysis of scientific literature [9]. Moreover, the application of AI algorithms has opened up new possibilities, enabling efficient data processing, pattern recognition, and knowledge extraction [10, 11]. Thus, by harnessing the power of AI, researchers can now delve into large-scale publication metrics, identify research trends, and track the influence and impact of scientific productions [10, 12, 13].

First, by leveraging natural language processing (NLP) algorithms, machine learning techniques, and deep learning approaches, AI can extract key information from scientific papers from a scientometric perspective to gain a comprehensive understanding of research trends, collaborations, and impact within specific domains [14].

Next, in terms of webometrics, AI algorithms can collect data from various online sources through web scraping, including web pages, blogs, forums, and social media posts. Machine learning, data mining algorithms, and deep learning (DL) techniques can extract data and patterns to help researchers understand and predict online users' behaviors, and digital impact [15, 16].

"Finally", through AI-powered algorithms, bibliometricians can analyze large-scale bibliographic and citation databases, such as Web of Science or Scopus, to uncover patterns, trends, and relationships among scientific productions [17].

These algorithms and approaches are helpful for policymakers and academicians to assess the impact of researchers, institutions, or scientific fields, facilitating evidence-based decisions, policy making, innovation mapping, and forecasting future-oriented developments [18].

While AI has shown great promise in improving the efficiency and accuracy of scientometric, webometric, and bibliometric analyses, there remains a lack of comprehensive understanding of the cutting-edge techniques and advancements in this rapidly evolving field. As researchers strive to harness the power of AI to gain deeper insights into scholarly communication patterns, citation networks, and the impact of research, it is crucial to conduct a systematic review that consolidates and synthesizes the latest developments and methodologies.

Therefore, the problem at hand is the absence of a comprehensive overview and analysis of the current state-of-the-art AI-enhanced techniques in scientometrics, webometrics, and bibliometrics. This knowledge gap inhibits researchers and practitioners from fully capitalizing on the potential benefits and advancements offered by AI in these domains. By conducting a systematic review, we aim to address this gap and provide a comprehensive understanding of the state-of-the-art AI techniques, their applications, and their impact on the field of informetrics.

In our study, we focus on these three specific fields (scientometrics, webometrics, and bibliometrics) because they represent key areas where the application of artificial intelligence (AI) has had a significant impact. AI techniques, such as machine learning and natural language processing, have greatly enhanced the analysis of large-scale bibliographic and web-based data, enabling more accurate and efficient measurement of scientific impact, knowledge diffusion, and web visibility.

Through this systematic review, we seek to shed light on the potential of AI to transform the way we measure and analyze scholarly communication, identify emerging research trends, and assess the impact of scientific publications. By doing so, we hope to inspire further research and innovation in the field of informetrics, ultimately leading to more accurate, efficient, and insightful analyses that can drive scientific progress and informed evidence-based decision-making.

**Materials and Methods**

Our study has involved conducting a thorough review of the existing literature to explore the various aspects and several indicators related to the use of AI-enhanced in Scientometrics, Webometrics, and Bibliometrics. Throughout the preparation of this manuscript, we have adhered to the PRISMA-ScR checklist and followed the recommended reporting guidelines for systematic reviews [19]. It is important to note that this manuscript has not been previously registered in PROSPERO or any similar database. We want to emphasize that while PROSPERO registration is typically associated with systematic reviews, we have made a deliberate decision not to register this specific review. This decision is based on the scope of our review, which does not strictly meet the eligibility criteria of PROSPERO, and the practicality within the limitations of our project. We want to assure readers that our literature search and selection process follow rigorous methodology, and our findings are reported transparently, thus in order to address any concerns regarding credibility.

*Research Questions*

1. "How do the cutting-edge techniques in AI-enhanced scientometrics contribute to the field of research evaluation and impact assessment?"
2. "What advancements have been made in AI-enhanced webometrics and how do they enhance the understanding of web-based information and online user behavior?"
3. "In what ways do the cutting-edge techniques in AI-enhanced bibliometrics revolutionize the analysis and measurement of scholarly publications and their impact?"
- Additionally, we are seeking answers to the following inquiries:

4. "What does the future hold for Scientometrics, Webometrics, and Bibliometrics with AI?"

5. "What are the ethical considerations that need to be taken into account when utilizing AI in Scientometrics, Webometrics, and Bibliometrics?"

*Inclusion and Exclusion Criteria*

During the study selection process, we implemented specific criteria to identify relevant articles from the database. We considered various types of articles, excluding systematic review articles as our aim is to concentrate on original research studies, and meta-analyses as they often have their own distinct inclusion and exclusion criteria that may differ from ours. The selected articles needed to focus on the use of AI to transform the measurement and analysis of scholarly communication, do identify emerging research trends and, evaluate the impact of scientific publications. Consequently, articles solely addressing the analysis of scholarly communication and the impact of scientific publications without any relevance to AI are excluded from the review. Through the application of these criteria, we ensure that the chosen studies directly address the analysis of AI-enhanced techniques in the field of scientometrics, webometrics, and bibliometrics, enabling us to provide a targeted and focused analysis for our research.

*Databases and Search Method*

We have conducted searches in several databases including ProQuest (LISTA & IBSS), EBSCO (LISTA), IEEE Explore, Web of Science, and Scopus to identify relevant studies. The search was limited to articles published between January 1, 2000, and September 2022, in order to encompass the most recent literature related to our research objectives. To ensure a comprehensive search strategy, we utilized a combination of broad search terms and conducted a nested search [20]. The search strategy involved using keywords that were relevant to our research topic, including variations and synonyms to maximize coverage. For instance, in Scopus, our search string included terms such as "AI" OR "Artificial Intelligence" AND "Scientometrics" OR "Webometrics" OR "Bibliometrics" or variations of it. By incorporating these keywords and using Boolean operators to combine them, our aim was to identify articles that focused on the impact, effectiveness, and evaluation of healthcare or smart health technologies. The specific search terms and string may have varied slightly for each database, but they followed a similar structure.

*Study selection*

The study selection process consisted of two steps to identify articles that met our inclusion criteria. Initially, two independent reviewers (HR.S. and E.H.) screened the titles and abstracts of the identified articles to determine their relevance to our research question and inclusion criteria. Any disagreements between the reviewers were resolved through discussion and consensus. If disagreements persisted, a third reviewer (M.A.) was consulted as an arbitrator. The third reviewer carefully examined the articles in question and provided input to reach a consensus. This approach ensured that the final selection of articles was (and is) based on collective agreement.

*Study quality appraisal*

The quality assessment of the included reviews was conducted by two researchers (HR.S. and E.H.) using the CASP Systematic Review Checklist (Appendix 1). We resolved any disagreements through discussion and reached a consensus on the quality of each study.

***Coding framework for analyzing the selected articles***

To ensure a systematic and consistent analysis of the selected articles, a coding framework was developed and applied. The coding framework consisted of several key categories and criteria that guided the analysis process. The following is an overview of the coding framework used:

1. *Category 1: Research Methodology*
   - Criteria: Identify the research methodology employed in each article (e.g., experimental, survey, case study, etc.).

2. *Category 2: AI Applications*
   - Criteria: Determine the specific applications of artificial intelligence discussed or utilized in each article (e.g., machine learning, natural language processing, data mining, etc.).

3. *Category 3: Metrics and Measures*
   - Criteria: Capture the different metrics and measures used or proposed in the articles for evaluating the impact or effectiveness of the AI applications in scientometrics, webometrics, and bibliometrics.

4. *Category 4: Ethical Considerations*
   - Criteria: Identify any ethical considerations or implications discussed in relation to the AI applications in the selected articles.

*5. Category 5: Future Implications*

- Criteria: Examine the discussions or predictions regarding the future implications and potential developments related to the use of AI in scientometrics, webometrics, and bibliometrics.

During the analysis, two independent researchers coded each article using this framework. Any discrepancies or disagreements in coding were resolved through discussion and consensus. Inter-coder reliability was assessed by calculating Cohen's kappa coefficient, which yielded a substantial agreement level of 0.85. By employing this coding framework, we aimed to provide a comprehensive analysis of the selected articles and ensure consistency in the evaluation of relevant aspects. The coding process allowed for a systematic examination of the research methodology, AI applications, metrics, ethical considerations, and future implications discussed in each article.

**Results**

The identification phase of our study involved conducting searches on various databases, such as ProQuest (LISTA & IBSS), EBSCO (LISTA), IEEE Explore, Web of Science, and Scopus. From these databases, a total of 1827 articles were initially found. After removing duplicate articles (962), we were left with 865 articles for further evaluation. By carefully assessing the exclusion criteria based on specific criteria, we excluded 356 articles, resulting in 509 articles for a more thorough analysis of their titles and abstracts. From a full-text analysis, we identified 61 of 121 articles that were relevant to our study's focus on AI in Scientometrics, Webometrics, and Bibliometrics. Therefore, our final dataset consisted of 61 articles, as shown in the flow chart provided in Figure 1.

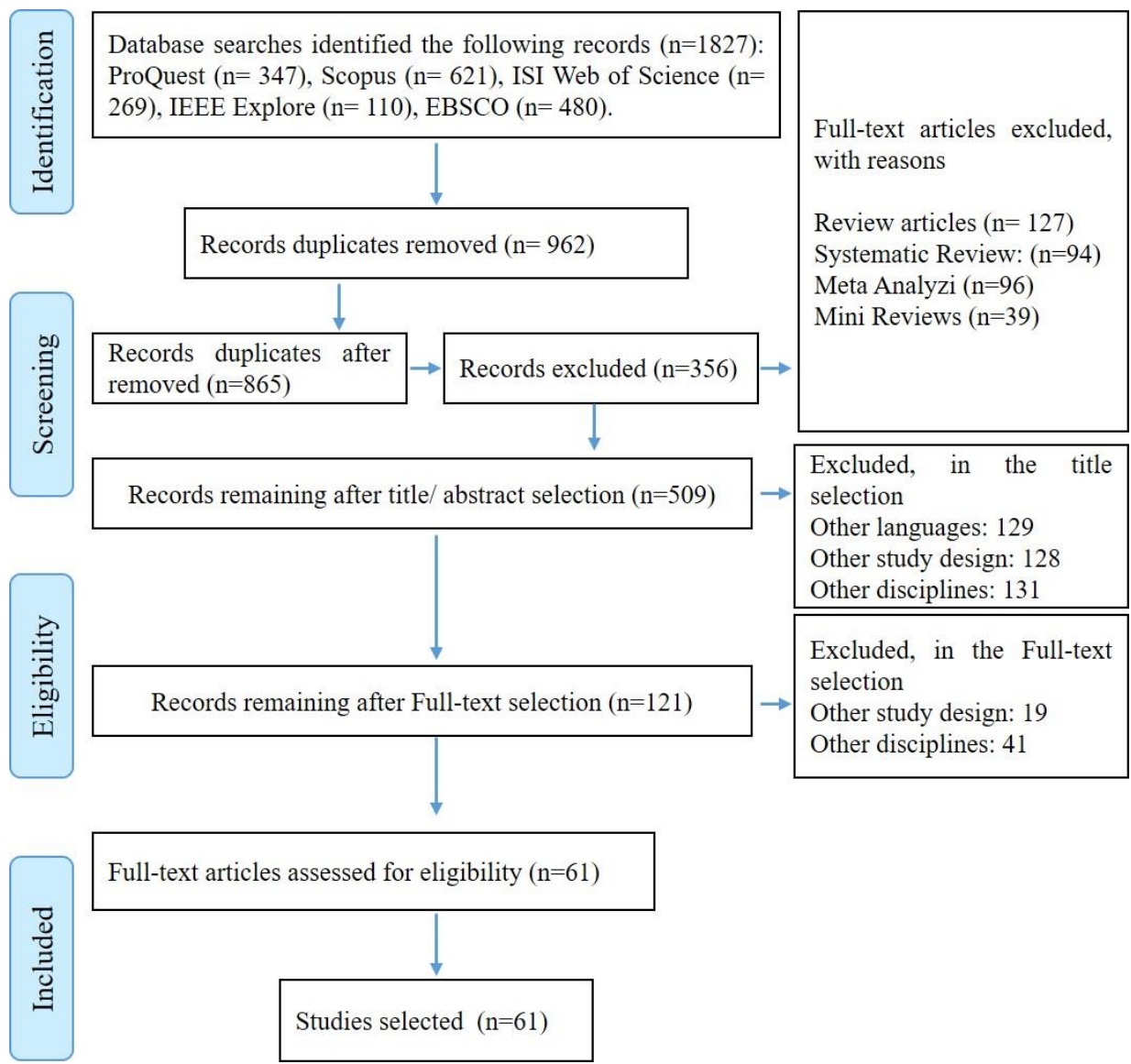

**Figure 1. Flowchart of the review of the articles, within PRISMA guidelines**

**RQ 1: AI and scientometrics**

In scientometrics, AI can provide several specific benefits including Publication Analysis, Citation Analysis, Prediction of Research Impact, Collaboration Analysis, Research Trend Analysis, and Knowledge Mapping. The AI benefits in such six subfields (Figure 2) have been discussed, e.g., in [21-31].

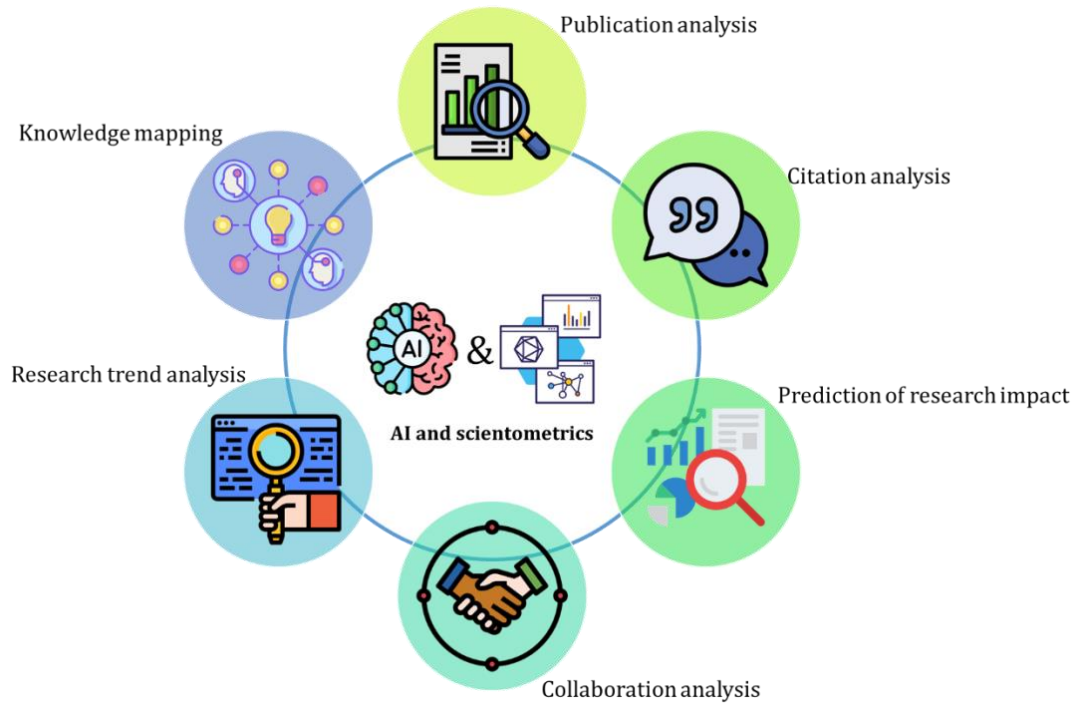

**Figure 2. Six specific benefits AI can provide to scientometrics; Source: by the authors**

These 12 studies demonstrate the potential benefits and strategies for utilizing AI capabilities in scientometrics. How AI can improve the quality, accessibility, and data collection processes in scientometric analyses is further highlighted in Table 1.

The main point is that AI algorithms can analyze large volumes of scientific publications and extract valuable information, such as author and co-author names, affiliations, keywords, and citations [21, 22]. As a result, researchers can gain insight into publication patterns, research networks, and collaborations within a particular scientific field [32, 33].

**Table 1. Studies Demonstrating the Usable Capacities of Artificial Intelligence for Scientometrics**

| References | Findings | Main Points |
|---|---|---|
| Donthu, Kumar et al. 2021, Rietz 2021, Caputo and Kargina 2022, Saeidnia, Kozak et al. 2023, Soleymani, Saeidnia et al. 2023 | *Publication Analysis* | AI can enhance the accuracy and efficiency of data collection and analysis in scientometrics. |
| Abrishami and Aliakbary | *Citation* | AI algorithms can effectively identify citation |

| 2019, Fazeli-Varzaneh, Ghorbi et al. 2021, Nicholson, Mordaunt et al. 2021, Caputo and Kargina 2022, Zhao and Feng 2022 | *Analysis* | patterns and analyze the impact of scientific publications. |
|---|---|---|
| Bertsimas, Brynjolfsson et al. , Kappen, van Klei et al. 2018, Moons, Wolff et al. 2019, Zhang and Wu 2020, Ma, Liu et al. 2021 | *Prediction of Research Impact* | AI can analyze large volumes of data to identify emerging research trends and predict future directions in scientometrics. |
| Mizoguchi, Imakura et al. 2022, Ullah, Shahid et al. 2022, Maghsoudi, Shokouhyar et al. 2023 | *Collaboration Analysis* | AI techniques can identify and analyze patterns of scientific collaborations, facilitating the understanding of collaborative networks in scientometrics. |
| Aparicio, Aparicio et al. 2019, Chen, Chen et al. 2020, Ceptureanu, Cerqueti et al. 2021, Jebari, Herrera-Viedma et al. 2021, Karpov, Pitsik et al. 2023 | *Research Trend Analysis* | AI can analyze scientific literature to identify emerging research areas and facilitate the discovery of new knowledge domains. |
| Corea and Corea 2019, Pedro, Subosa et al. 2019, Pasquinelli and Joler 2021, Liang, Luo et al. 2022 | *Knowledge Mapping* | knowledge mapping via AI-based metrics is to map a visual representation of the knowledge and information within an organization or domain. This mapping can help identify gaps in knowledge and potential opportunities for innovation. Knowledge mapping can also help organizations make more informed decisions. |

Moreover, AI algorithms can analyze citation networks to identify the impact and influence of scientific papers, as well as the relationships between different research works [22, 24, 31]. Researchers can use this method to identify highly cited and influential papers, - even sleeping beauties [34] as well as to understand the dynamics of scientific knowledge dissemination.

Interestingly, AI techniques can be employed to predict the impact of scientific research based on various factors, such as author reputation, journal quality, and citation patterns [27]. Analyzing historical data allows AI models to provide insights into the potential impact of research, enabling researchers and institutions to determine the best course of action.

Co-authorship networks can be analyzed by AI to identify and understand research collaborations [28, 30]. By analyzing publication history, author affiliations, and co-authorship patterns, AI can help researchers identify potential collaborators and research networks, enabling better collaboration and knowledge exchange.

In order to identify emerging research trends and topics, AI can analyze large-scale scientific literature [23, 26, 35] For example, by utilizing natural language processing techniques, AI algorithms can automatically extract keywords, topics, and trends from scientific publications, helping researchers identify new research directions and stay up-to-date with the latest advancements in their field.

"Finally", AI can map the scientific knowledge landscape by analyzing the relationships between different scientific papers, keywords, and concepts [25, 29]. In addition to facilitating literature reviews, hypothesis generation, and research planning, this visualization allows researchers to visualize and understand the structure and evolution of knowledge within a specific research area.

**RQ 2: AI and webometrics**

In webometrics, AI can provide several specific benefits including Web Crawling and Data Collection, Web Link Analysis, Web Content Analysis, Social Media Analysis, Web Impact Analysis, and Recommender Systems as sketched in Figure 3, and e.g. it has been demonstrated through papers like [9, 10, 21, 36-45].

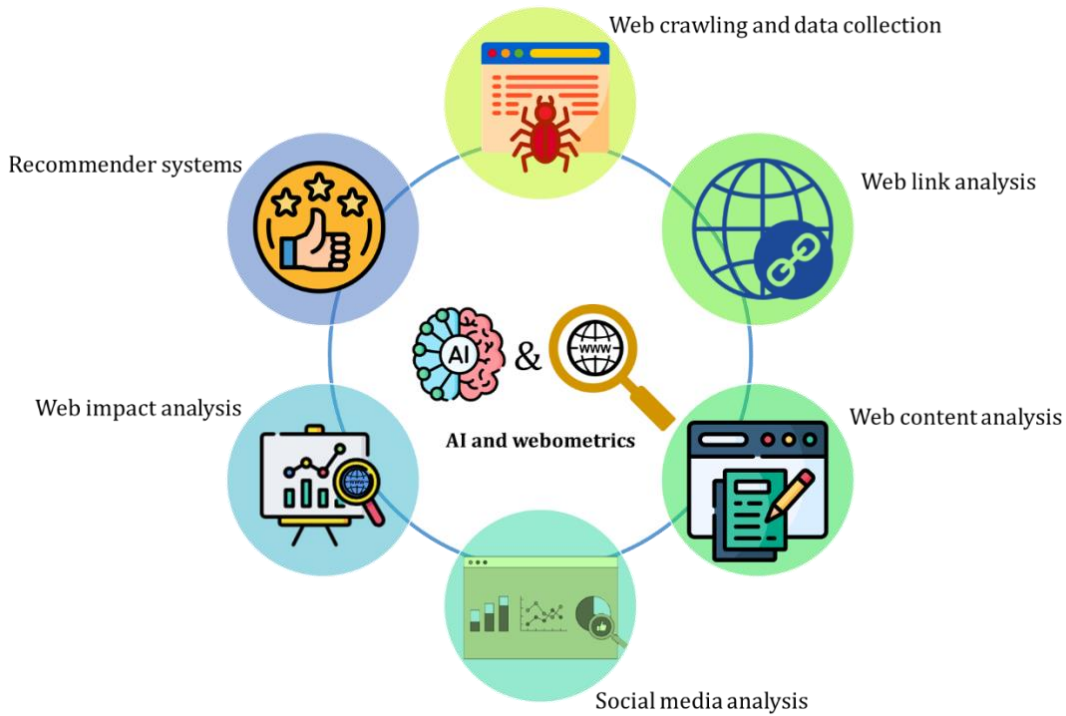

**Figure 3. Six specific benefits AI can provide to webometrics; Source: by the authors**

These 6 considerations point to the potential benefits and suggest focussed strategies for utilizing AI capabilities in webometrics. The resulting findings highlight how AI can improve the quality, accessibility, and data collection processes in webometrics analyses, as outlined in Table 2.

Indeed, algorithms based on artificial intelligence can automatically crawl and collect data from websites, including institutional websites, scientific research portals, and online repositories [39, 42]. This enables researchers to gather large amounts of web-based information for analysis, including publication data, author profiles, and citation patterns.

In order to understand the relationship between publications, websites, and authors, artificial intelligence approaches can analyze hyperlink structures and web link patterns [9, 43]. By analyzing the link structure, AI algorithms can identify influential websites and authors, as well as detect communities, collaborations, and research networks within the web-based scientific ecosystem [17].

**Table 2. Studies Demonstrating the Usable Capacities of Artificial Intelligence for Webometrics**

| References | Findings | Main Points |
|---|---|---|
| Serrano 2018, Brewer, Westlake et al. 2021, Khder 2021, Alaidi, Roa'a et al. 2022 | *Web Crawling and Data Collection* | AI can automate the web crawling process, extracting data from websites and improving the efficiency of data collection for webometrics. |
| Yuan, Chen et al. 2018, Thomas and Mathur 2019, Korkmaz, Sahingoz et al. 2020, Wang and Yu 2021, Xu, Liu et al. 2021 | *Web Link Analysis* | AI algorithms can effectively analyze web link structures and identify influential websites or pages. |
| Yuan, Chen et al. 2018, Serafini and Reid 2019, Dutta 2021, Kiesel, Meyer et al. 2021, Maulud, Zeebaree et al. 2021, Jalil, Usman et al. 2023 | *Web Content Analysis* | AI can automate the analysis of web content, extracting relevant information and identifying trends or patterns. |
| Balaji, Annavarapu et al. 2021, Amjad, Younas et al. 2022, Grover, Kar et al. 2022, Wu, Dodoo et al. 2022 | *Social Media Analysis* | AI can enhance the accuracy and efficiency of data collection and analysis in webometrics. To assess the use of AI in web impact assessment. AI-based metrics can provide more comprehensive and accurate measures of web impact, considering various factors beyond traditional link counts. |
| Thomas and Mathur 2019, Barclay, Taylor et al. 2021, Xu, Liu et al. 2021 | *Web Impact Analysis* | AI-based metrics can provide more comprehensive and accurate measures of web impact, considering various factors beyond traditional link counts. |
| Ceptureanu, Cerqueti et al. 2021, Zhang, Lu et al. 2021 | *Recommender Systems* | AI algorithms can analyze web usage data to understand user behavior, preferences, and trends, aiding in the improvement of web design and user experience. |

AI techniques, such as natural language processing and machine learning, can be employed to analyze the content of webpages and scientific publications available online [40, 41]. This enables researchers to extract key information, such as keywords, topics, and sentiments, from web-based documents, facilitating comprehensive analysis and understanding of research outputs.

AI can analyze social media platforms, such as Twitter, to understand the online discussions, trends, and interactions related to scientific research [36, 38, 44]. By analyzing hashtags, mentions, and user behavior, AI algorithms can identify influential research topics, key opinion leaders, and potential collaborations within the online scientific community, as demonstrated in such previous works.

AI can assess the impact and visibility of scientific research on the web [37, 46]. Indeed, by analyzing web traffic, page views, and social media metrics, AI algorithms can provide insights into the online visibility, dissemination, and engagement of scientific publications, authors, and research institutions.

"Finally", AI-powered recommender systems can assist researchers in discovering relevant scientific websites, online resources, and research collaborations [35, 45]. These papers, based on user preferences, reading behavior, and web usage data, show that personalized recommendations can be generated using AI algorithms, making it easier for researchers to explore the web-based scientific landscape and discover new opportunities for further research.

**RQ 3: AI and bibliometrics**

In bibliometrics, AI can provide several specific benefits including Automated Data Collection, Citation Analysis, Author Disambiguation, Co-authorship Analysis, Research Impact Analysis, Text Mining, and Recommender Systems (see Figure 4) as analyzed in [28-30, 47-53].

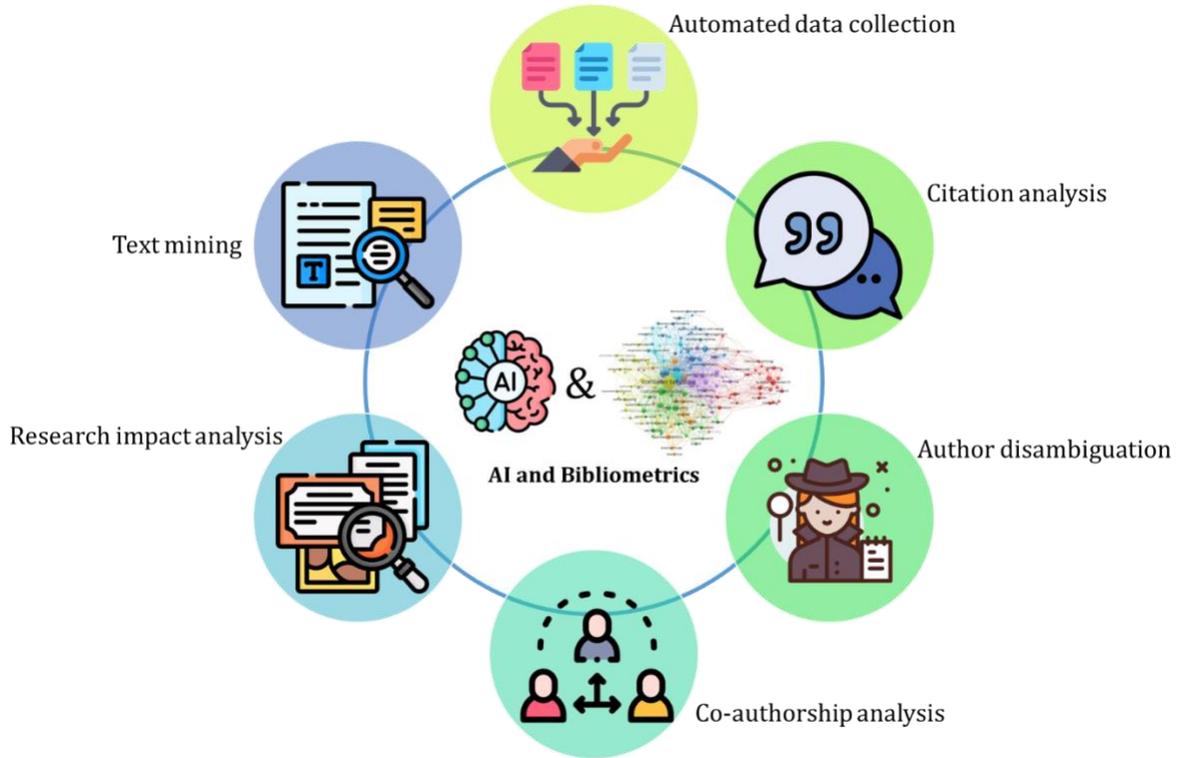

**Figure 4. Six specific benefits AI can provide to bibliometrics. Source: by the authors**

These 6 studies demonstrate the potential benefits and strategies for utilizing AI capabilities in bibliometrics. They highlight how AI can improve the quality, accessibility, and data collection processes in bibliometrics analyses (Table 3), among the main outcome points.

**Table 3. Studies Demonstrating the Usable Capacities of Artificial Intelligence for Bibliometrics**

| References | Findings | Main Points |
|---|---|---|
| Cox and Mazumdar , Donthu, Kumar et al. 2021, Rietz 2021, Caputo and Kargina 2022, Saeidnia, Kozak et al. 2023, Soleymani, Saeidnia et al. 2023 | *Automated Data Collection* | AI can enhance the accuracy and efficiency of data collection and analysis in bibliometrics. |
| Abrishami and Aliakbary 2019, Fazeli-Varzaneh, Ghorbi et al. 2021, Nicholson, Mordaunt et al. 2021, Caputo and Kargina 2022, Zhao and Feng 2022, Maghsoudi, Shokouhyar et al. 2023 | *Citation Analysis* | AI can analyze citation patterns and identify influential publications, aiding in the evaluation of scholarly impact. |
| Tekles and Bornmann 2020, Mihaljević and | *Author | AI algorithms can accurately |

| Santamaría 2021, Rehs 2021, Abramo and D'Angelo 2023 | *Disambiguation* | disambiguate author names, improving the reliability of bibliometric analyses. |
| --- | --- | --- |
| Fonseca Bde, Sampaio et al. 2016, Grodzinski, Grodzinski et al. 2021, Ullah, Shahid et al. 2022, Maghsoudi, Shokouhyar et al. 2023 | *Co-authorship Analysis* | AI techniques can identify patterns of co-authorship and analyze collaboration networks in bibliometrics. |
| Donthu, Kumar et al. 2021, Al-Jamimi, BinMakhashen et al. 2022, Loan, Nasreen et al. 2022 | *Research Impact Analysis* | AI can predict the potential impact of research articles, assisting in identifying influential publications. |
| Mrowinski, Fronczak et al. 2017, Eisenstein 2019, Kang, Cai et al. 2020, Mohammadzadeh, Ausloos et al. 2023, Saeidnia and Lund 2023 | *Text Mining* | AI algorithms can automatically extract relevant keywords from scholarly articles, improving information retrieval and analysis. |

It has been shown that AI algorithms can automatically collect bibliographic data from various sources, such as online databases, academic libraries, and digital repositories [21, 49]. This saves time and effort for researchers involved in data collection, allowing them to focus on other aspects of bibliometric analysis.

Thought-provokingly, AI can analyze citation networks to identify influential papers, authors, and journals [28, 31], - as already mentioned in the section "AI and Scientometrics". By examining citation patterns and relationships, AI algorithms can help researchers understand the impact and visibility of research outputs, as well as identify key research trends and collaborations.

Interestingly, AI techniques can be employed to disambiguate authors with similar names, a common issue in bibliometrics [47, 53]. By analyzing author affiliations, publication history, and co-authorship networks, in order to ensure the accuracy of bibliometric analyses, AI algorithms can effectively identify authors with similar names and distinguish them from one another.

As also already mentioned in the section "AI and Scientometrics", through AI one can analyze co-authorship networks to identify collaborations and research networks [28, 30]. By examining co-authorship patterns and relationships, AI algorithms can help researchers understand the dynamics and structure of collaborations, as well as identify influential researchers and research teams. This can also be an advantage at funding time.

Easily, AI can analyze bibliometric indicators, such as citation counts and h-index, to assess the impact and visibility of individual researchers, research groups, and institutions [21, 48, 52]. In so doing, AI algorithms can provide insights into research productivity, citation patterns, and research impact over time, assisting researchers and institutions in assessing research fame or performance.

Last but not least, AI techniques, including natural language processing, can be utilized to analyze the textual content of research publications [50, 51]. In this manner, keywords, topics, and sentiments can be extracted from the literature, - also mentioning plagiarism control [54, 55], thereby facilitating comprehensive analysis and understanding of research findings [56].

**Discussion**

The above hopefully rather complete, at least extensive, literature survey allows a critical assessment of the state of AI in informatics science.

First, the findings in Table 1 have significant implications for scientometrics. They highlight the potential benefits and strategies for utilizing artificial intelligence (AI) capabilities in scientometrics analyses. The mentioned studies clearly demonstrate that AI can enhance the accuracy and efficiency of data collection and analysis in scientometrics [21, 22, 32, 33]. By automating various tasks, AI algorithms can reduce human errors and biases, ensuring more reliable and consistent results. This enhanced accuracy and efficiency save time and resources, allowing researchers to focus on high-level analyses and interpretations.

AI-based citation analysis methods, author disambiguation techniques, and predictive models showcased in the mentioned studies provide researchers with powerful tools for improving data collection and analysis in scientometrics [22, 24, 31, 34]. AI algorithms can effectively identify citation patterns, analyze the impact of scientific publications, and predict research trends. These capabilities enable researchers to gain deeper insights into the scientific landscape and make informed decisions.

Traditional citation counts have limitations in measuring research impact. However, the studies demonstrate that AI-based metrics can provide more comprehensive and accurate measures of research impact [25, 29]. By considering various factors beyond citations, such as social media mentions, downloads, and collaborations, AI algorithms can provide a more holistic view of the impact of scientific publications.

AI techniques showcased in the studies can analyze scientific literature to identify emerging research areas and patterns of scientific collaborations [28, 30]. This enables researchers to stay updated with the latest trends, discover new knowledge domains, and foster collaborations with relevant stakeholders.

AI-based peer review systems, as highlighted in one of the studies, can enhance the efficiency and objectivity of the peer review process [27, 57]. By automating parts of the review process, AI can ensure the publication of high-quality research, reduce biases, and provide faster feedback to authors. This improves the overall quality of scientometrics analyses and accelerates the dissemination of scientific knowledge.

Another study demonstrates that AI can assist in detecting instances of scientific misconduct, such as plagiarism and data fabrication [55]. By analyzing large volumes of data and comparing it against established standards, AI algorithms can identify potential cases of misconduct, ensuring the integrity of scientometrics analyses [17, 54, 55].

In summary, the findings in Table 1 demonstrate that AI has the potential to revolutionize techniques and approaches of scientometrics. AI capabilities improve the accuracy, efficiency, and reliability of data collection, analysis, and assessment of research impact. They enable the identification of emerging research areas, collaboration networks, and instances of scientific misconduct. Ultimately, these findings contribute to the advancement of scientometrics research, improving the quality, accessibility, and overall understanding of the scientific landscape.

Table 2 presents studies that demonstrate the potential benefits and strategies for utilizing artificial intelligence (AI) capabilities in webometrics [9, 36-44, 46, 58-66]. The findings in this table have significant implications for webometrics, as they highlight how AI can enhance various aspects of the field.

Indeed, the studies mentioned in Table 2 showcase that AI can improve data collection and analysis in webometrics, and how. In particular, AI algorithms can automate the process of gathering web data, such as web links, page content, and user behavior. This automation not only saves time and effort but also ensures the collection of larger and more diverse datasets, leading to more comprehensive webometric analyses.

AI techniques, such as machine learning and network analysis, are employed in the studies to improve web link analysis in webometrics [9, 43]. These techniques enable researchers to identify influential websites, web pages, and online communities [42, 59]. AI algorithms can analyze the structure and dynamics of web links, providing insights into the connectivity and impact of web resources [39, 42, 58, 59].

AI algorithms can analyze web content to extract relevant information and identify trends in webometrics [41, 62, 64, 65]. Natural language processing (NLP) techniques can be employed to automatically extract keywords, topics, and sentiments from web pages [40, 41, 62-65]. This automated analysis enhances the efficiency and accuracy of webometric studies, enabling researchers to gain insights into web-based information dissemination and trends [40, 41].

AI-based metrics and algorithms can provide advanced web impact assessment in webometrics [46, 60]. Beyond traditional link counts, AI algorithms can consider factors such as user behavior, social media mentions, and content engagement to measure the impact of web resources [37, 46, 60]. This comprehensive assessment helps researchers and organizations understand the reach and influence of web content [37, 46].

Web usage mining refers to the analysis of user behavior on the web. AI techniques, such as machine learning and data mining, can be employed to analyze user interactions, navigation paths, and preferences on websites. This analysis helps researchers understand user behavior patterns, improve web design, and enhance user experience.

AI algorithms can improve the efficiency and effectiveness of web crawling and data extraction in webometrics. These algorithms can automatically navigate through web pages, extract relevant data, and filter out irrelevant or duplicate information. This automation streamlines the data collection process, enabling researchers to analyze larger volumes of web data.

In a nutshell, let it be mentioned that the findings in Table 2 demonstrate that AI has the potential to significantly enhance webometrics. By improving data collection, web link analysis, content analysis, impact assessment, web usage mining, and data extraction, AI algorithms empower researchers to conduct more comprehensive and accurate webometric analyses. These advancements contribute to a deeper understanding of web-based information dissemination, user behavior, and the impact of web resources.

Thirdly, Table 3 presents studies that demonstrate the potential benefits and strategies for utilizing artificial intelligence (AI) capabilities in bibliometrics [21, 22, 24, 28, 30-34, 47-51, 53-56, 67-72]. The findings in this table have significant implications for bibliometrics, as they highlight how AI can enhance various aspects of the field.

AI algorithms can improve publication analysis in bibliometrics [21, 22, 32, 33, 67]. By automatically extracting metadata from scientific publications, such as author names, affiliations, citations, and keywords, AI techniques can streamline the data collection process and improve accuracy [21, 22, 32, 33, 49, 67]. This automation allows researchers to analyze larger volumes of publications, facilitating comprehensive bibliometric analyses [21, 22, 32, 33, 67].

AI techniques can enhance citation analysis in bibliometrics. AI algorithms can automatically identify and analyze citation patterns, such as co-citation and bibliographic coupling [22, 24, 28, 31, 34, 68]. These algorithms can also identify citation networks and clusters, providing insights into the relationships among scientific publications [22, 24, 28, 31, 34, 68]. This analysis helps researchers understand the influence and impact of scholarly work [24, 31, 34].

AI algorithms can aid in author disambiguation, a critical task in bibliometrics [28, 30, 70-72]. By analyzing various factors, such as author names, affiliations, and publication history, AI techniques can accurately identify and disambiguate authors with similar names [30, 72]. This disambiguation ensures accurate attribution of scholarly work and improves the reliability of bibliometric analyses [28, 30, 71, 72].

AI techniques, such as machine learning and data mining, can be employed to develop predictive models in bibliometrics [50, 51, 55, 56]. These models can forecast future publication trends, identify emerging research areas, and predict research impact [50, 51, 54, 55]. By analyzing patterns and relationships in large bibliographic datasets, AI algorithms can provide valuable insights into the future direction of scientific research [54-56].

AI algorithms can analyze collaboration networks among researchers in bibliometrics. By analyzing co-authorship patterns, affiliations, and research collaborations, AI techniques can identify influential researchers, research groups, and institutions. This analysis not only helps researchers understand the dynamics of collaboration but should be expected to foster interdisciplinary research, beside more usual links.

AI techniques can enhance research evaluation in bibliometrics. By considering various factors beyond traditional citation counts, such as social media mentions, downloads, and media coverage, AI algorithms can provide more comprehensive metrics for evaluating research impact. This comprehensive evaluation helps researchers, institutions, and funding agencies make informed decisions and allocate resources effectively.

Furthermore, improved or specifically written AI algorithms can assist in detecting instances of scientific misconduct, and prove plagiarism and data fabrication.

Concisely, the findings in Table 3 demonstrate that AI has the potential to significantly enhance bibliometrics. By improving publication analysis, citation analysis, author disambiguation, predictive models, collaboration analysis, and research evaluation, AI algorithms empower researchers to conduct more comprehensive and accurate bibliometric analyses. These advancements contribute to a deeper understanding of scholarly communication, research impact, and collaboration dynamics in the scientific community.

**RQ 4: Future of Scientometrics, Webometrics, and Bibliometrics with AI**

From the above, one can imagine if not research gaps, at least directions for further progress. Artificial intelligence (AI) has the potential to significantly benefit all three fields - scientometrics, webometrics, and bibliometrics. However, the extent to which AI can perform and its future implications may vary in each field.

It has been shown here above that AI can greatly enhance scientometrics by improving data collection and analysis, text mining and information retrieval, identification of emerging research trends, visualization techniques, research evaluation, and collaboration and networking. The use of AI algorithms can automate processes, increase efficiency, and provide deeper insights into scientific literature [21-31]. The future of scientometrics with AI is likely to involve more advanced AI algorithms, improved integration of various data sources, and increased automation, leading to more accurate and comprehensive analyses.

AI can play a significant role in webometrics by improving data collection and analysis, web link analysis, web content analysis, web impact assessment, web usage mining, and efficient web crawling and data extraction [9, 10, 21, 36-41, 43-45]. AI techniques can help extract valuable information from the web, analyze user behavior, and assess the impact of web resources [9, 36-45]. The future of webometrics with AI may involve advancements in AI algorithms for web data analysis, better understanding of user behavior, and improved techniques for web impact assessment.

AI can enhance bibliometrics by improving publication analysis, citation analysis, author disambiguation, predictive models, collaboration analysis, and research evaluation. AI algorithms can automate processes, provide accurate citation analysis, and develop predictive models for future research trends [28-30, 47-53]. The future of bibliometrics with AI may involve more advanced techniques for author disambiguation, improved prediction models, integration of alternative metrics, and better evaluation of research impact beyond traditional citation counts.

In terms of which field AI can perform the most, it is difficult to determine a clear winner. AI has the potential to significantly benefit all three fields and can perform exceptionally well in each, depending on the specific applications and techniques employed. The effectiveness of AI in each field will also depend on the availability and quality of data, the complexity of the analysis required, and the specific research questions being addressed.

The future of these three areas with AI is promising. As AI technologies continue to advance, we can expect more sophisticated algorithms, improved integration of various data sources, and enhanced automation and efficiency in scientometrics, webometrics, and bibliometrics. The use of AI will likely lead to more accurate and comprehensive analyses, better understanding of research trends and impact, and improved decision-making processes in academia, research institutions, and funding agencies.

**RQ 5: Ethical Considerations of Scientometrics, Webometrics, and Bibliometrics with AI**

Beside positive aspects so outlined, nevertheless, the use of artificial intelligence (AI) in scientometrics, webometrics, and bibliometrics raises important ethical considerations that should be carefully addressed.

AI algorithms often require access to large amounts of data, including personal and sensitive information [73]. It is crucial to ensure that proper data protection measures are in place to safeguard privacy and prevent unauthorized access [74]. Data anonymization and encryption techniques should be employed, and compliance with relevant data protection regulations should be followed [75].

AI algorithms can be prone willingly or inadvertently, to bias, whence can result in unfair or discriminatory outcomes [17, 76]. It is important to ensure that AI models are trained on diverse and representative datasets to avoid perpetuating existing biases [77]. Regular monitoring and auditing of AI systems should be conducted to identify and address any biases that may arise [78].

Sometimes, AI algorithms can be complex and opaque, making it difficult to understand how they arrive at their decisions [79]. Thus, it is important to promote transparency and explainability in AI models used in scientometrics, webometrics, and bibliometrics. Researchers and users should have access to information about the data used, the algorithms employed, and the decision-making processes of the AI systems [76, 79].

As AI systems become more autonomous, it is essential to establish clear lines of accountability and responsibility [80]. Developers, researchers, and users should be aware of their roles and responsibilities in ensuring the responsible and ethical use of AI in these fields. This includes addressing any potential biases, errors, or unintended consequences that may arise from the use of AI.

In cases where personal data is involved, obtaining informed consent from individuals is crucial [78]. Researchers and organizations should have robust consent management processes in place to ensure that individuals understand how their data will be used and have the ability to provide or withdraw consent.

Furthermore, the use of AI in scientometrics, webometrics, and bibliometrics may have implications for employment and society as a whole. It is important to consider the potential impact on jobs, the distribution of resources, and the broader societal implications. Measures should be taken to mitigate any negative effects and ensure a fair and equitable transition. Regular monitoring and evaluation of AI systems should be conducted to assess their performance, identify any biases or ethical concerns, and make necessary improvements. This ongoing monitoring and evaluation process should involve interdisciplinary collaboration and engagement with stakeholders.

Addressing these ethical considerations requires a multidisciplinary approach involving researchers, policymakers, ethicists, and stakeholders from various fields. Open dialogue, transparency, and ongoing evaluation are essential to ensure that AI is used responsibly and ethically in scientometrics, webometrics, and bibliometrics.

**Conclusion**

In this report, we emphasize the importance and potential of integrating AI algorithms with scientometrics, webometrics, and bibliometrics, through numerous examples in the literature.

The paradigm shift undergone by AI algorithms in these fields has been shown to have revealed new possibilities for analysis, prediction, and pattern mining-based recommendations. Within this review, the paper contributes to underscoring the prominent prospects and value of integrating AI in scientometrics, webometrics, and bibliometrics, i.e., to signify the synergy that can be achieved and fostered through this integration.

In brief, AI helps scientometrics by providing efficient and accurate methods to analyze and derive insights from scientific publications, citation networks, and collaborative relationships. This should enable researchers to gain a deeper understanding of scientific knowledge, trends, and impact, facilitating better decision-making and advances in scientific research. Moreover, AI enhances webometrics by providing efficient and automated methods to analyze web-based scientific data, understand link structures and social interactions, assess web impact, and provide personalized recommendations. This enables researchers to gain insights into the web-based scientific ecosystem, facilitate collaborations, and improve research visibility and impact in the digital age. In addition, AI enhances the bibliometrics field of activities by automating data collection, providing accurate author disambiguation, analyzing citation networks, assessing research impact, and providing personalized recommendations. This enables researchers to gain insights into scholarly communication, assess research performance, and make informed decisions in their bibliometric analyses. Overall, AI presents an efficient and scalable approach to scientometrics, webometrics, and bibliometrics, enabling researchers to extract meaningful insights from vast and diverse sources of scientific information.

In conclusion, the integration of artificial intelligence (AI) into scientometrics, webometrics, and bibliometrics holds significant potential for advancing research and understanding in these fields. AI can enhance data collection, analysis, prediction, and evaluation processes, providing researchers with valuable insights and improving decision-making processes.

However, the use of AI in these areas also raises important ethical considerations that must be carefully addressed. Data privacy and security, bias, and fairness, transparency and explainability, accountability and responsibility, informed consent, impact on employment and society, and continuous monitoring and evaluation are among the key ethical considerations that should be taken into account. To ensure the responsible and ethical use of AI, interdisciplinary collaboration, stakeholder engagement, and ongoing evaluation are crucial. Researchers, policymakers, ethicists, and stakeholders from various fields should work together to develop guidelines, frameworks, and best practices that promote ethical AI use in scientometrics, webometrics, and bibliometrics. By addressing these ethical considerations, we can harness the full potential of AI to advance knowledge, improve research practices, and contribute to the betterment of society while ensuring fairness, transparency, and accountability in the use of these technologies.

**Limitations**

In this particular study, we did not include the gray literature in our search and review process, nor did we manually search in Google Scholar. Instead, our intention was to focus on searching in reliable databases. While Google Scholar is often referred to as a database, it is actually a search engine that may not include high-quality articles and may only retrieve reliable studies. By not searching in Google Scholar, we aimed to minimize the number of overlapping studies.

However, it is important to note that this highly technical approach may have resulted in overlooking certain articles, which could regretfully lead to our study excluding relevant information. We consider that up to the time of writing and submitting this paper, we safeguard against much omission. Yet, for future studies, it may be beneficial to conduct a comprehensive review that includes the gray literature, in order to provide readers with a broader perspective.

**References**

1. Darko A, Chan AP, Adabre MA, Edwards DJ, Hosseini MR, Ameyaw EE. Artificial intelligence in the AEC industry: Scientometric analysis and visualization of research activities. Automation in Construction. 2020;112:103081.
2. Park S, Park HW. A webometric network analysis of electronic word of mouth (eWOM) characteristics and machine learning approach to consumer comments during a crisis. Profesional de la Información. 2020;29(5).
3. Van Raan A. Scientometrics: State-of-the-art. Scientometrics. 1997;38(1):205-18.
4. Bharvi D, Garg K, Bali A. Scientometrics of the international journal Scientometrics. Scientometrics. 2003;56(1):81-93.
5. Thelwall M, Vaughan L, Björneborn L. Webometrics. Annual review of information science and technology. 2005;39(1):81-135.
6. Björneborn L, Ingwersen P. Perspective of webometrics. Scientometrics. 2001;50:65-82.
7. McBurney MK, Novak PL, editors. What is bibliometrics and why should you care? Proceedings IEEE international professional communication conference; 2002: IEEE.
8. Cooper ID. Bibliometrics basics. Journal of the Medical Library Association: JMLA. 2015;103(4):217.
9. Xu Y, Liu X, Cao X, Huang C, Liu E, Qian S, et al. Artificial intelligence: A powerful paradigm for scientific research. The Innovation. 2021;2(4).

10.	Melnikova E. Big data technology in the set of methods and means of scientific research in modern scientometrics. Scientific and Technical Information Processing. 2022;49(2):102-7.
11.	Tapeh ATG, Naser M. Artificial intelligence, machine learning, and deep learning in structural engineering: a scientometrics review of trends and best practices. Archives of Computational Methods in Engineering. 2023;30(1):115-59.
12.	Saeidnia H. Using ChatGPT as a Digital/Smart Reference Robot: How May ChatGPT Impact Digital Reference Services? Information Matters. 2023;2(5).
13.	Saeidnia H. Open AI, ChatGPT: To Be, or Not to Be, That Is the Question. Information Matters. 2023;3(6).
14.	Yuan S, Shao Z, Wei X, Tang J, Hall W, Wang Y, et al. Science behind AI: The evolution of trend, mobility, and collaboration. Scientometrics. 2020;124:993-1013.
15.	Chaudhuri N, Gupta G, Vamsi V, Bose I. On the platform but will they buy? Predicting customers' purchase behavior using deep learning. Decision Support Systems. 2021;149:113622.
16.	G. Martín A, Fernández-Isabel A, Martín de Diego I, Beltrán M. A survey for user behavior analysis based on machine learning techniques: current models and applications. Applied Intelligence. 2021;51(8):6029-55.
17.	Saeidnia HR. Ethical artificial intelligence (AI): confronting bias and discrimination in the library and information industry. Library Hi Tech News. 2023;ahead-of-print(ahead-of-print). doi: 10.1108/LHTN-10-2023-0182.
18.	Hain D, Jurowetzki R, Lee S, Zhou Y. Machine learning and artificial intelligence for science, technology, innovation mapping and forecasting: Review, synthesis, and applications. Scientometrics. 2023;128(3):1465-72.
19.	Tricco AC, Lillie E, Zarin W, O'Brien KK, Colquhoun H, Levac D, et al. PRISMA Extension for Scoping Reviews (PRISMA-ScR): Checklist and Explanation. Annals of internal medicine. 2018;169(7):467-73. Epub 2018/09/05. doi: 10.7326/m18-0850. PubMed PMID: 30178033.
20.	Holzmann GJ, Peled DA, Yannakakis M. On nested depth first search. The Spin Verification System. 1996;32:81-9.
21.	Donthu N, Kumar S, Mukherjee D, Pandey N, Lim WM. How to conduct a bibliometric analysis: An overview and guidelines. Journal of business research. 2021;133:285-96.
22.	Caputo A, Kargina M. A user-friendly method to merge Scopus and Web of Science data during bibliometric analysis. Journal of Marketing Analytics. 2022;10(1):82-8.
23.	Chen X, Chen J, Cheng G, Gong T. Topics and trends in artificial intelligence assisted human brain research. PLoS One. 2020;15(4):e0231192.
24.	Abrishami A, Aliakbary S. Predicting citation counts based on deep neural network learning techniques. Journal of Informetrics. 2019;13(2):485-99.
25.	Corea F, Corea F. AI knowledge map: How to classify AI technologies. An introduction to data: Everything you need to know about AI, big data and data science. 2019:25-9.
26.	Jebari C, Herrera-Viedma E, Cobo MJ. The use of citation context to detect the evolution of research topics: a large-scale analysis. Scientometrics. 2021;126(4):2971-89.
27.	Ma A, Liu Y, Xu X, Dong T. A deep-learning based citation count prediction model with paper metadata semantic features. Scientometrics. 2021;126(8):6803-23.
28.	Maghsoudi M, Shokouhyar S, Ataei A, Ahmadi S, Shokoohyar S. Co-authorship network analysis of AI applications in sustainable supply chains: Key players and themes. Journal of cleaner production. 2023;422:138472.
29.	Pedro F, Subosa M, Rivas A, Valverde P. Artificial intelligence in education: Challenges and opportunities for sustainable development. 2019.
30.	Ullah M, Shahid A, Roman M, Assam M, Fayaz M, Ghadi Y, et al. Analyzing interdisciplinary research using Co-authorship networks. Complexity. 2022;2022.

31. Zhao Q, Feng X. Utilizing citation network structure to predict paper citation counts: A Deep learning approach. Journal of Informetrics. 2022;16(1):101235.
32. Saeidnia HR, Kozak M, Lund B, Mannuru NR, Keshavarz H, Elango B, et al. Design, Development, Implementation, and Evaluation of a Mobile Application for Academic Library Services: A Study in a Developing Country. Information Technology and Libraries. 2023;42(3).
33. Soleymani H, Saeidnia HR, Ausloos M, Hassanzadeh M. Selective dissemination of information (SDI) in the age of artificial intelligence (AI). Library Hi Tech News. 2023;ahead-of-print(ahead-of-print). doi: 10.1108/LHTN-08-2023-0156.
34. Fazeli-Varzaneh M, Ghorbi A, Ausloos M, Sallinger E, Vahdati S. Sleeping beauties of coronavirus research. Ieee Access. 2021;9:21192-205.
35. Ceptureanu S, Cerqueti R, Alexandru A, Popescu D, Dhesi G, Ceptureanu E. Influence of blockchain adoption on technology transfer, performance and supply chain integration, exibility and responsiveness. A case study from IT&C medium size enterprises. Studies in Informatics and Control. 2021;30(3):61-74.
36. Amjad S, Younas M, Anwar M, Shaheen Q, Shiraz M, Gani A. Data mining techniques to analyze the impact of social media on academic performance of high school students. Wireless Communications and Mobile Computing. 2022;2022:1-11.
37. Barclay I, Taylor H, Preece A, Taylor I, Verma D, de Mel G. A framework for fostering transparency in shared artificial intelligence models by increasing visibility of contributions. Concurrency and Computation: Practice and Experience. 2021;33(19):e6129.
38. Grover P, Kar AK, Dwivedi YK. Understanding artificial intelligence adoption in operations management: insights from the review of academic literature and social media discussions. Annals of Operations Research. 2022;308(1-2):177-213.
39. Khder MA. Web Scraping or Web Crawling: State of Art, Techniques, Approaches and Application. International Journal of Advances in Soft Computing & Its Applications. 2021;13(3).
40. Maulud DH, Zeebaree SR, Jacksi K, Sadeeq MAM, Sharif KH. State of art for semantic analysis of natural language processing. Qubahan academic journal. 2021;1(2):21-8.
41. Serafini F, Reid SF. Multimodal content analysis: expanding analytical approaches to content analysis. Visual Communication. 2019:1470357219864133.
42. Serrano W. Neural networks in big data and Web search. Data. 2018;4(1):7.
43. Wang W, Yu L. UCrawler: a learning-based web crawler using a URL knowledge base. Journal of Computational Methods in Sciences and Engineering. 2021;21(2):461-74.
44. Wu L, Dodoo NA, Wen TJ, Ke L. Understanding Twitter conversations about artificial intelligence in advertising based on natural language processing. International Journal of Advertising. 2022;41(4):685-702.
45. Zhang Q, Lu J, Jin Y. Artificial intelligence in recommender systems. Complex & Intelligent Systems. 2021;7(1):439-57. doi: 10.1007/s40747-020-00212-w.
46. Xu Y, Liu X, Cao X, Huang C, Liu E, Qian S, et al. Artificial intelligence: A powerful paradigm for scientific research. The Innovation. 2021;2(4):100179. doi: https://doi.org/10.1016/j.xinn.2021.100179.
47. Abramo G, D'Angelo CA. How reliable are unsupervised author disambiguation algorithms in the assessment of research organization performance? Quantitative Science Studies. 2023:1-26.
48. Al-Jamimi HA, BinMakhashen GM, Bornmann L. Use of bibliometrics for research evaluation in emerging markets economies: a review and discussion of bibliometric indicators. Scientometrics. 2022;127(10):5879-930.
49. Cox AM, Mazumdar S. Defining artificial intelligence for librarians. Journal of Librarianship and Information Science.0(0):09610006221142029. doi: 10.1177/09610006221142029.
50. Eisenstein J. Introduction to natural language processing: MIT press; 2019.

51. Kang Y, Cai Z, Tan C-W, Huang Q, Liu H. Natural language processing (NLP) in management research: A literature review. Journal of Management Analytics. 2020;7(2):139-72.
52. Loan FA, Nasreen N, Bashir B. Do authors play fair or manipulate Google Scholar h-index? Library Hi Tech. 2022;40(3):676-84.
53. Rehs A. A supervised machine learning approach to author disambiguation in the Web of Science. Journal of Informetrics. 2021;15(3):101166.
54. Mohammadzadeh Z, Ausloos M, Saeidnia HR. ChatGPT: high-tech plagiarism awaits academic publishing green light. Non-fungible token (NFT) can be a way out. Library Hi Tech News. 2023.
55. Saeidnia HR, Lund BD. Non-fungible tokens (NFT): a safe and effective way to prevent plagiarism in scientific publishing. Library Hi Tech News. 2023;40(2):18-9.
56. Mrowinski MJ, Fronczak P, Fronczak A, Ausloos M, Nedic O. Artificial intelligence in peer review: How can evolutionary computation support journal editors? PloS one. 2017;12(9):e0184711.
57. Piva F, Tartari F, Giulietti M, Aiello MM, Cheng L, Lopez-Beltran A, et al. Predicting future cancer burden in the United States by artificial neural networks. Future Oncology. 2020;17(2):159-68.
58. Brewer R, Westlake B, Hart T, Arauza O. The Ethics of Web Crawling and Web Scraping in Cybercrime Research: Navigating Issues of Consent, Privacy, and Other Potential Harms Associated with Automated Data Collection. In: Lavorgna A, Holt TJ, editors. Researching Cybercrimes: Methodologies, Ethics, and Critical Approaches. Cham: Springer International Publishing; 2021. p. 435-56.
59. Alaidi AHM, Roa'a M, ALRikabi H, Aljazaery IA, Abbood SH. Dark web illegal activities crawling and classifying using data mining techniques. iJIM. 2022;16(10):123.
60. Thomas DM, Mathur S, editors. Data analysis by web scraping using python. 2019 3rd International conference on Electronics, Communication and Aerospace Technology (ICECA); 2019: IEEE.
61. Korkmaz M, Sahingoz OK, Diri B, editors. Detection of phishing websites by using machine learning-based URL analysis. 2020 11th International Conference on Computing, Communication and Networking Technologies (ICCCNT); 2020: IEEE.
62. Yuan H, Chen X, Li Y, Yang Z, Liu W, editors. Detecting phishing websites and targets based on URLs and webpage links. 2018 24th International Conference on Pattern Recognition (ICPR); 2018: IEEE.
63. Dutta AK. Detecting phishing websites using machine learning technique. PloS one. 2021;16(10):e0258361.
64. Jalil S, Usman M, Fong A. Highly accurate phishing URL detection based on machine learning. Journal of Ambient Intelligence and Humanized Computing. 2023;14(7):9233-51.
65. Kiesel J, Meyer L, Kneist F, Stein B, Potthast M, editors. An empirical comparison of web page segmentation algorithms. European Conference on Information Retrieval; 2021: Springer.
66. Balaji T, Annavarapu CSR, Bablani A. Machine learning algorithms for social media analysis: A survey. Computer Science Review. 2021;40:100395.
67. Rietz T. Designing AI-based systems for qualitative data collection and analysis. 2021.
68. Nicholson JM, Mordaunt M, Lopez P, Uppala A, Rosati D, Rodrigues NP, et al. Scite: A smart citation index that displays the context of citations and classifies their intent using deep learning. Quantitative Science Studies. 2021;2(3):882-98.
69. Mihaljević H, Santamaría L. Disambiguation of author entities in ADS using supervised learning and graph theory methods. Scientometrics. 2021;126(5):3893-917. doi: 10.1007/s11192-021-03951-w.
70. Tekles A, Bornmann L. Author name disambiguation of bibliometric data: A comparison of several unsupervised approaches1. Quantitative Science Studies. 2020;1(4):1510-28. doi: 10.1162/qss_a_00081.
71. Grodzinski N, Grodzinski B, Davies BM. Can co-authorship networks be used to predict author research impact? A machine-learning based analysis within the field of degenerative cervical myelopathy research. Plos one. 2021;16(9):e0256997.

72. Fonseca Bde P, Sampaio RB, Fonseca MV, Zicker F. Co-authorship network analysis in health research: method and potential use. Health research policy and systems. 2016;14(1):34. Epub 2016/05/04. doi: 10.1186/s12961-016-0104-5. PubMed PMID: 27138279; PubMed Central PMCID: PMCPMC4852432.
73. Hancock JT, Naaman M, Levy K. AI-mediated communication: Definition, research agenda, and ethical considerations. Journal of Computer-Mediated Communication. 2020;25(1):89-100.
74. Safdar NM, Banja JD, Meltzer CC. Ethical considerations in artificial intelligence. European journal of radiology. 2020;122:108768.
75. Pratomo AB, Mokodenseho S, Aziz AM. Data Encryption and Anonymization Techniques for Enhanced Information System Security and Privacy. West Science Information System and Technology. 2023;1(01):1-9.
76. Ferrer X, van Nuenen T, Such JM, Coté M, Criado N. Bias and discrimination in AI: a cross-disciplinary perspective. IEEE Technology and Society Magazine. 2021;40(2):72-80.
77. Ferrara E. Fairness And Bias in Artificial Intelligence: A Brief Survey of Sources, Impacts, And Mitigation Strategies. arXiv preprint arXiv:230407683. 2023.
78. Gichoya JW, Thomas K, Celi LA, Safdar N, Banerjee I, Banja JD, et al. AI pitfalls and what not to do: mitigating bias in AI. The British Journal of Radiology. 2023:20230023.
79. von Eschenbach WJ. Transparency and the black box problem: Why we do not trust AI. Philosophy & Technology. 2021;34(4):1607-22.
80. Yazdanpanah V, Gerding EH, Stein S, Dastani M, Jonker CM, Norman TJ, et al. Reasoning about responsibility in autonomous systems: challenges and opportunities. AI & SOCIETY. 2023;38(4):1453-64.